\documentclass[aps,prd,twocolumn,showpacs,notitlepage,eqsecnum,
superscriptaddress]{revtex4-1}

\usepackage[centertags]{amsmath}
\usepackage{amssymb}
\usepackage{latexsym}
\usepackage{enumerate}
\usepackage{graphicx}
\usepackage{mathrsfs}
\usepackage{hyperref}
\usepackage{stmaryrd}
\usepackage{import}
\usepackage{tensor}
\usepackage{color}
\usepackage{bm}
\usepackage{multirow}
\usepackage{breakurl}
\usepackage{float}

\newcommand{\bi}{\begin{itemize}}
\newcommand{\ei}{\end{itemize}}
\newcommand{\be}{\begin{equation}}
\newcommand{\ee}{\end{equation}}

\renewcommand{\l}{\left(}
\renewcommand{\r}{\right)}
\renewcommand{\a}{\alpha}

\renewcommand{\O}{\Omega}
\renewcommand{\o}{\omega}
\renewcommand{\th}{\theta}

\newcommand{\q}{\quad}

\newcommand{\vp}{\varphi}
\newcommand{\ti}{\tilde}

\newcommand{\pa}{\partial}

\newcommand{\bscal}[1]{\boldsymbol{\mathcal{#1}}}

\begin{document}

\title{Fast spectral source integration in black hole perturbation calculations}

\author{Seth Hopper}
\affiliation{School of Mathematical Sciences and Complex \& Adaptive Systems 
Laboratory, University College Dublin, Belfield, Dublin 4, Ireland}
\author{Erik Forseth}
\author{Thomas Osburn}
\author{Charles R. Evans}
\affiliation{Department of Physics and Astronomy, University of North 
Carolina, Chapel Hill, North Carolina 27599, USA}

\begin{abstract}
This paper presents a new technique for achieving spectral accuracy and 
fast computational performance in a class of black hole perturbation and 
gravitational self-force calculations involving extreme mass ratios and 
generic orbits.  Called \emph{spectral source integration} (SSI), this 
method should see widespread future use in problems that entail (i) 
point-particle description of the small compact object, (ii) frequency 
domain decomposition, and (iii) use of the background eccentric geodesic 
motion.  Frequency domain approaches are widely used in both perturbation 
theory flux-balance calculations and in local gravitational self-force 
calculations.  Recent self-force calculations in Lorenz gauge, using the 
frequency domain and method of extended homogeneous solutions, have been 
able to accurately reach eccentricities as high as $e \simeq 0.7$.  We 
show here SSI successfully applied to Lorenz gauge.  In a double precision 
Lorenz gauge code, SSI enhances the accuracy of results and makes a factor 
of three improvement in the overall speed.  The primary initial application of 
SSI--for us its \emph{raison d'\^{e}tre}--is in an arbitrary precision 
\emph{Mathematica} code that computes perturbations of eccentric orbits 
in the Regge-Wheeler gauge to extraordinarily high accuracy (e.g., 200 
decimal places).  These high accuracy eccentric orbit calculations would not 
be possible without the exponential convergence of SSI.  We believe the 
method will extend to work for inspirals on Kerr, and will be the subject of 
a later publication.  SSI borrows concepts from discrete-time signal 
processing and is used to calculate the mode normalization coefficients 
in perturbation theory via sums over modest numbers of points around an 
orbit.  A variant of the idea is used to obtain spectral accuracy in 
solution of the geodesic orbital motion.

\end{abstract}


\maketitle

\section{Introduction}
\label{sec:intro}

Merging compact binaries are a promising source of detectable gravitational 
waves.  Accurate theoretical models serve as templates to assist detection 
and will aid in estimating an event's physical parameters.  Three 
complementary theoretical approaches exist \cite{Leti14} for computing 
relativistic binaries: numerical relativity \cite{BaumShap10,LehnPret14}, 
post-Newtonian (PN) theory \cite{Will11,Blan14}, and gravitational 
self-force (GSF) and black hole perturbation (BHP) calculations 
\cite{DrasHugh05,Bara09,PoisPounVega11,Thor11,Leti14}.  The effective-one-body 
(EOB) formalism, drawing calibration of its parameters from the above 
approaches, then provides a synthesis \cite{BuonDamo99,BuonETC09,Damo10,
HindETC13,Damo13,TaraETC14}.

The GSF approach assumes the existence of, and exploits, a small ratio 
$q =\mu/M \ll 1$ between the component masses.  The field and motion of 
the smaller body are calculated in a perturbation expansion in powers of $q$ 
\cite{MinoSasaTana97,QuinWald97}.  Though restricted to small $q$, the GSF 
is valid throughout the strong field regime.  GSF/BHP calculations are 
most relevant to potential future eLISA observations of extreme-mass-ratio 
inspirals (EMRIs) $q \simeq 10^{-7}$-$10^{-4}$ \cite{AmarETC14} but might 
pertain to Advanced LIGO observations if there exists a fortuitous population 
of intermediate-mass-ratio inspirals (IMRIs) $q \simeq 10^{-3}$-$10^{-2}$ 
\cite{BrowETC07,AmarETC07}.  The dominant approach to the GSF treats 
the small body as a point mass \cite{PoisPounVega11}, then calculates the 
metric perturbation and the local self-force using mode-sum regularization 
\cite{BaraOri00}.  Calculations are done directly in the time domain (TD) 
\cite{Mart04,BaraSago07,FielHestLau09,CaniSopu14} or via decomposition into 
Fourier-harmonic modes in the frequency domain (FD) 
\cite{Detw08,HoppEvan10,Akca11,AkcaWarbBara13,OsbuETC14}.  Alternative means 
of calculating the GSF include effective source calculations 
\cite{VegaDetw08,VegaWardDien11,WardETC12} and direct Green function 
calculations \cite{CasaETC09,CasaETC13,WardETC14}.  

The PN approach has no restriction on $q$ but is most accurate for wide, 
low frequency orbits.  Just as the GSF, PN, and NR approaches separately 
inform EOB, there has been considerable activity in recent years in making 
comparisons between GSF/BHP and PN theory \cite{Detw08,SagoBaraDetw08,
BlanETC09,BlanETC10}, including calculations 
at very high accuracies \cite{Fuji12,ShahFrieWhit14,Shah14,ShahPoun15}.  
These high precision calculations, until recently all done for circular 
orbits, utilize the analytic function expansion formalism of Mano, Suzuki, 
and Takasugi (MST) \cite{ManoSuzuTaka96b} and make use of arbitrary precision 
coding.  

Several of us were intrigued by the idea of extending MST calculations, and 
these comparisons, to include eccentric orbits.  Initial results of that now 
successful effort will be described elsewhere \cite{ForsEvanHopp15}, but the 
project led to the necessary development of the technique reported here.  
Modeling EMRIs with large eccentricities is essential, since astrophysical 
considerations suggest \cite{AmarETC07,AmarETC14} they have a distribution 
peaked about $e \simeq 0.7$ \cite{HopmAlex05} as they enter eLISA's 
passband.  Advanced LIGO inspirals are expected to have nearly circular 
orbits, but small eccentricity corrections may be important \cite{Moor15}. 

The spectral technique described here benefits FD calculations of the 
``geodesic self-force'' (i.e., first-order perturbations derived using 
geodesics of the background geometry) in E/IMRIs with eccentric orbits.  
In eccentric-orbit FD calculations the Fourier transform spreads the 
influence of the point particle source across a range of radii.  Mode by mode, 
the resulting source functions are integrated against a Green function 
over this radial libration region, a procedure that has been followed for 
decades \cite{CutlKennPois94} for BHP problems.  FD calculations of eccentric 
orbit GSF became feasible after Barack, Ori, and Sago \cite{BaraOriSago08} 
found the method of extended homogeneous solutions (EHS), thus allowing 
Fourier synthesis at the particle location without encountering Gibbs 
behavior.  Originally demonstrated for scalar models 
\cite{BaraOriSago08} and early-on extended to master equations in the 
Regge-Wheeler-Zerilli (RWZ) formalism \cite{HoppEvan10}, EHS has recently been 
applied to coupled systems in Lorenz gauge \cite{AkcaWarbBara13,OsbuETC14}.  

To those familiar with EHS, the new method can be outlined briefly here.  
See Secs.~\ref{sec:method} and \ref{sec:methodSystem} for details.  Here we 
couch the discussion in terms of the RWZ case (Sec.~\ref{sec:SSI}), where EHS 
entails calculating arbitrarily normalized causal homogeneous solutions 
($\hat{X}_{lmn}^{-}$, $\hat{X}_{lmn}^{+}$) and integrating them in 
product with stress-tensor projections over the source region.  The result 
is a set of normalization coefficients $C_{lmn}^{\pm}$ that encode the 
orbital motion's imprint in the field perturbation.  Then extended homogeneous 
solutions are assembled in the TD, and subsequently abutted at the 
instantaneous particle location.

The new technique provides a means of calculating the $C_{lmn}^{\pm}$ (or 
their equivalent in other gauges) with spectral accuracy.  The integral for 
the normalization coefficients is typically manipulated 
\cite{BaraOriSago08,HoppEvan10,WarbBara11} into a form like
\begin{align}
C_{lmn}^\pm  
&= \frac{1}{W_{lmn} T_r} \int_0^{T_r}
\bar{E}_{lmn}^{\pm}(t) \ e^{i n\O_r t} \ dt ,
 \label{eqn:Cold}
\end{align}
where $\bar{E}_{lmn}^{\pm}(t)$ is a periodic function of the radial motion 
derived from spherical harmonic projection of the point source and 
integration over the $\hat{X}_{lmn}^{\mp}(r)$.  Details are found in 
Sec.~\ref{sec:method} but for the nonce it is enough to say that computing 
\eqref{eqn:Cold} is difficult to do with ODE or numerical quadrature 
integrators at high accuracies beyond double precision and is impossible to 
do at extraordinarily high accuracies like $100$ or more decimal places.  
The new method, called \emph{spectral source integration} (SSI), replaces 
the integral with a remarkably simple sum 
\begin{align}
\label{eqn:Cnew}
C_{lmn}^\pm  
&= \frac{1}{N W_{lmn}} \sum_{k=0}^{N-1} \bar{E}^{\pm}_{lmn}(t_k) 
\ e^{i n \O_r t_k } ,
\end{align}
which involves merely sampling the source function $\bar{E}_{lmn}^{\pm}(t)$ 
at a modest number $N$ of equally-spaced points around the closed radial 
motion.  This sum converges \emph{exponentially} with increases in $N$.  

The FD approach with use of Fourier series (FS) has been a part of BHP theory 
for decades.  The FS and normalization coefficients converge exponentially 
with $n$, allowing the FS to be truncated.  The new method makes a crucial 
use of that standard approximation, recognizing that truncation of the FS 
representation (of e.g., a source term) generates a \emph{bandlimited 
function}.  That in turn invokes the machinery of the Nyquist-Shannon 
sampling theorem.  The truncated FS can itself be replaced by discrete 
equally-spaced sampling of the TD function.  Then, discrete sampling and 
periodicity allow a discrete function of finite length $N$ to serve as an 
accurate TD representation.  Furthermore, the finite discrete function is 
dual to a discrete Fourier transform (DFT) spectrum, computable with an 
FFT.  The DFT spectrum is an approximation, between its Nyquist frequencies, 
of the original FS spectrum, but can be made exponentially accurate with 
increases in $N$.  It is then possible to replace integrals like 
\eqref{eqn:Cold} with finite sums like \eqref{eqn:Cnew} and achieve spectral 
convergence there too.  In essence, SSI provides a \emph{completion} of the 
FD approach by bringing to bear concepts in discrete-time signal processing.

This paper shows application of SSI to FD BHP and geodesic GSF calculations 
of eccentric Schwarzschild E/IMRIs in both RWZ and Lorenz gauges.  
We also demonstrate in Sec.~\ref{sec:orbit} that a related approach provides 
arbitrarily accurate solutions of the geodesic equations themselves.  SSI 
may be applicable to Kerr BHP \cite{DrasHugh05,DrasHugh06,FujiHikiTago09} 
and GSF calculations, the subject of an upcoming paper.  In addition, SSI 
has the potential to benefit the Green function approach to GSF calculations 
\cite{WardETC14}.  

This paper is organized as follows.  Sec.~\ref{sec:orbit} considers the 
orbital problem.  In Sec.~\ref{sec:orbitReview} we review bound eccentric 
geodesic motion about a Schwarzschild black hole and set the notation.  
Sec.~\ref{sec:orbitSpec} describes the spectral approach for integrating 
the orbit equations with geometric convergence, and shows numerical results. 
Appendix \ref{sec:appendix} gives a simple analytic calculation of the 
exponential fall-off in Fourier coefficients in part of the orbital problem.  
Next the SSI method is described in Sec.~\ref{sec:method} through its 
application in the RWZ formalism to provide spectral solution of master 
equations.  A brief review of how the RWZ problem is solved in the FD using 
EHS is given in Sec.~\ref{sec:RWZEHS}.  Then Sec.~\ref{sec:SSI} lays out 
the SSI method, the heart of this paper, and displays a set of numerical 
results.  We discuss some related findings in the numerical analysis 
literature in Sec.~\ref{sec:SSItrap}.  Having shown SSI applied to a single 
perturbation equation, we present next in Sec.~\ref{sec:methodSystem} its 
application in Lorenz gauge, demonstrating that the method allows systems 
of equations to be solved with spectral convergence.  Our conclusions are 
drawn in Sec.~\ref{sec:conclusions}.

In this paper we set $G = c = 1$ and use the metric signature $+2$.

\section{Spectral integration of bound orbital motion}
\label{sec:orbit}

The new method is first applied to solving the equations of bound geodesic 
motion.  This proves to be a necessary first step to using SSI to solve the 
first-order perturbation equations when working at accuracies well beyond 
double precision.  At double precision, it leads to a more efficient 
computation of the orbit.  We consider geodesic motion about a 
Schwarzschild black hole in this paper.  Application of SSI to general 
orbits about a Kerr black hole will be taken up in a subsequent 
paper.  We begin with a brief review of the problem and notation.

\subsection{Geodesic motion and the relativistic anomaly}
\label{sec:orbitReview}

We consider generic bound motion of a small mass $\mu$, taken to be a point 
particle, around a Schwarzschild black hole of mass $M$ in the test body 
(geodesic) limit $\mu/M \rightarrow 0$.  Schwarzschild coordinates 
$x^{\mu} = (t,r,\theta, \varphi )$ are used, with the line element having 
the form
\be
ds^2 = -f dt^2 + f^{-1} dr^2
+ r^2 \left( d\theta^2 + \sin^2\theta \, d\varphi^2 \right) ,
\ee
where $f(r) = 1 - 2M/r$.  

Let the worldline of the particle be described by the functions
$x_p^{\a}(\tau)=\left[t_p(\tau),r_p(\tau),\th_p(\tau),\varphi_p(\tau)\right]$
of proper time $\tau$ (or some other convenient curve parameter).  Subscript 
$p$ indicates location of the particle.  The four-velocity is 
$u^{\alpha} = dx_p^{\alpha}/d\tau$.  Without loss of generality the motion 
is confined to the equatorial plane, $\theta_p(\tau)=\pi/2$. 

The orbit is parametrized in terms of the (dimensionless) semi-latus rectum 
$p$ and the eccentricity $e$ (see \cite{CutlKennPois94,BaraSago10}).  These 
constants are related to the usual constant specific energy 
$\mathcal{E} = -u_t$ and specific angular momentum 
$\mathcal{L} = u_{\varphi}$.  Additionally, pericentric $r_{\rm min}$ and 
apocentric $r_{\rm max}$ radii are introduced, which are related to $p$ and 
$e$ by the following equations 
\be
\label{eqn:defeandp}
p = \frac{2 r_{\rm max} r_{\rm min}}{M (r_{\rm max} + r_{\rm min})} , 
\q \q
e = \frac{r_{\rm max} - r_{\rm min}}{r_{\rm max} + r_{\rm min}},
\ee
\be
r_{\rm max} = \frac{pM}{1-e}, 
\q \q
r_{\rm min} = \frac{pM}{1+e} .
\ee
Bound eccentric orbits satisfy $\mathcal{E} < 1$ and 
$\mathcal{L} > 2 \sqrt{3} M$.  These in turn imply $p \ge 6 + 2 e$, with 
the boundary of stable orbits $p = 6 + 2 e$ being the separatrix 
\cite{CutlKennPois94}.

As is usual, $\tau$ is replaced as the curve parameter by Darwin's 
relativistic anomaly $\chi$, in terms of which the radial position is given 
a Keplerian-appearing form \cite{Darw61}  
\be
r_p \l \chi \r = \frac{pM}{1+ e \cos \chi} .
\ee
The equations for the remaining functions take the form
\begin{align}
\label{eqn:tDarwin}
\frac{dt_p}{d \chi} &= \frac{r_p \l \chi \r^2}{M (p - 2 - 2 e \cos \chi)}
 \sqrt{\frac{(p-2)^2 -4 e^2}{p -6 -2 e \cos \chi} } ,
\\
\label{eqn:tauDarwin}
\frac{d\tau_p}{d \chi} &= \frac{M p^{3/2}}{(1 + e \cos \chi)^2} 
\sqrt{ \frac{p - 3 - e^2}{p - 6 - 2 e \cos \chi} } ,
\\
\label{eqn:phiDarwin}
\frac{d \varphi_p}{d\chi} 
&= \sqrt{ \frac{p}{p - 6 - 2 e \cos \chi} } .
\end{align}
The last equation, describing azimuthal motion, has an analytic solution 
\be
\vp_p(\chi) = \sqrt{\frac{4 p}{p - 6 - 2 e}} \, \,
F\left(\frac{\chi}{2} \, \middle| \, -\frac{4 e}{p - 6 - 2 e}  \right) ,
\ee
where $F(x|m)$ is the incomplete elliptic integral of the first kind
\cite{GradETC07}.  The other two equations are typically solved numerically.

To solve \eqref{eqn:tDarwin} and \eqref{eqn:tauDarwin}, each equation can be 
regarded as either a numerical quadrature or an initial value problem (IVP) 
\cite{PresETC93}.  Cutler, Kennefick, and Poisson \cite{CutlKennPois94} took 
the former approach and used Romberg's method.  In the more complicated 
Kerr geodesic problem, Drasco and Hughes \cite{DrasHugh06} initially solved 
for the motion using a numerical quadrature routine but later switched to 
use of a quasi-analytic approach developed by Fujita and Hikida \cite{FujiHiki09}.  (Indeed, this 
quasi-analytic method involving rapid evaluation of elliptic integrals stands 
as a third route to solution.)  In more recent work 
\cite{BaraSago10,HoppEvan10,AkcaWarbBara13,OsbuETC14}, 
Eqns.~\eqref{eqn:tDarwin} and \eqref{eqn:tauDarwin} have simply been 
integrated using Runge-Kutta routines.  At double precision the 
distinction is trivial and errors in the orbit are of minimal concern.  
Recently, however, several of us have turned attention \cite{ForsEvanHopp15} 
to making extraordinarily high precision (e.g., 200 decimal place) BHP and 
GSF calculations for eccentric EMRIs using the MST formalism 
\cite{ManoSuzuTaka96b} (henceforth the MST code).  It proved 
necessary to develop a new means of efficiently calculating the orbit to 
arbitrary precision, as well as doing the same for the perturbation source 
integration (Secs.~\ref{sec:SSI} and \ref{sec:SSIlorenz}). 

The MST code is written in \emph{Mathematica} to make use of its arbitrary 
precision functionality.  Initially, we used its NDSolve function to compute 
orbits but found such integrations became prohibitively expensive for 
errors of order $\lesssim 10^{-40}$.  The alternative approach we found 
turns out to be a simple application of the SSI concepts.  In fact, the 
arguments laid out in the next two subsections are key to understanding the 
SSI development.  Shortly, we will discuss solving \eqref{eqn:tDarwin} to 
obtain $t_p(\chi)$ (integration of \eqref{eqn:tauDarwin} follows in like 
fashion).  But first we address some general considerations.

\subsection{Spectral integration: general considerations}
\label{sec:SpecInt}

Let $dI/d\chi = g(\chi)$ with $g(\chi)$ (the source) being both a periodic and 
a smooth function.  We are interested in integrating $g$ to find $I(\chi)$.  
We can assume $g(\chi)$ is complex, but in orbital motion applications 
the functions will be real.  The periodicity of $g$ suggests utilizing a FS 
expansion and then calculating the integral for $I(\chi)$ term by term.  At 
first glance this approach is not very helpful since, even if we truncate 
the FS, the expression for $I(\chi)$ would require computing a large number 
of definite integrals numerically for the FS coefficients.  Fortunately, the 
smoothness of $g(\chi)$ helps in several ways.  In many cases, the FS
amplitudes $\ti{\mathcal{G}}_n$ will fall in magnitude exponentially (shown 
numerically for orbital motion presently; see also Appendix A).  Even in 
calculations with hundreds of decimal places of accuracy, the FS can then 
be truncated to a modest number of terms.  At whatever adopted level of 
accuracy, replacing $g(\chi)$ with a truncated FS introduces an approximation 
that is \emph{bandlimited}.  

We then recall that bandlimited signals play a key role in the Nyquist-Shannon 
sampling theorem: a function that contains only frequencies $f$ with 
$|f| \le B$ is completely determined by its discrete (equally-spaced) samples 
(in this case in $\chi$) occurring at the Nyquist rate $2 B$ (i.e., with 
spacing $\Delta\chi = \tfrac{1}{2} B^{-1}$).  If we combine discrete sampling 
with the periodicity of radial motion, then only a finite total number $N$ of 
samples in $\chi$ need be considered.  We replace $g(\chi)$ again--this time 
with its finite sampling $g_k = g(\chi_k) = g(k \Delta\chi)$, where 
$k = 0,\ldots , N-1$.  This new representation of the source has its own 
DFT spectrum $\mathcal{G}_n$ (with $n = -N/2,\ldots , N/2-1$), which can be 
computed with the FFT algorithm \cite{PresETC93}.  In contrast to the FS 
spectrum $\ti{\mathcal{G}}_n$, the DFT spectrum $\mathcal{G}_n$ exhibits a 
periodicity of its own, $\mathcal{G}_{n+jN} = \mathcal{G}_n$, for arbitrary 
integer $j$.  However, aliasing can be avoided if the DFT spectrum is only 
used at the $N$ frequencies within its Nyquist bounds.  Then for an accuracy 
goal that is sufficiently high (i.e., high enough $N$, found iteratively), 
the DFT spectrum $\mathcal{G}_n$ is virtually indistinguishable from the 
FS spectrum $\ti{\mathcal{G}}_n$.  Using the DFT representation, it is then 
possible to compute $g(\chi)$ at any location either via Fourier 
interpolation or using the Whittaker cardinal function \cite{Sten81} on the 
circle (i.e., convolution with the Dirichlet kernel).  Furthermore, the 
source can be integrated or differentiated term by term to accuracies 
comparable to the initial goal.

To summarize:
\bi
\item
The (perhaps complex) function $g(\chi)$ is periodic and $C^{\infty}$.  
\item
It can be represented as a FS with spectrum 
$\ti{\mathcal{G}}_{n}$ with $n \rightarrow \pm\infty$.  
\item
The FS spectrum can be truncated to some $n_{\rm min} \le n \le n_{\rm max}$ 
subject to an accuracy goal.  
\item
The approximate but very accurate truncated FS is a bandlimited 
function.
\item
The Nyquist-Shannon sampling theorem implies the truncated FS representation 
can itself be replaced in the TD with discrete sampling.  
\item
Sampling plus periodicity implies a discrete representation of finite 
length $N$. 
\item
Finite sampling representation in the TD implies one-to-one correspondence 
via the DFT with a FD periodic spectrum $\mathcal{G}_{n}$.  
\item 
The DFT spectrum within the Nyquist range approximates well the original 
FS spectrum if $N$ is sufficiently large, allowing 
$\ti{\mathcal{G}}_{n} \rightarrow \mathcal{G}_{n}$.  
\item
The DFT representation in the TD can be integrated and interpolated to 
spectral accuracy.
\ei

\subsection{Spectral solution of the orbital motion}
\label{sec:orbitSpec}

In practice, the orbit equations \eqref{eqn:tDarwin} and \eqref{eqn:tauDarwin} 
have source functions that are real and even.  Hence we can represent them
with a discrete cosine transform (DCT) \cite{AhmeNataRao74}.  In turn the 
integral for $t_p(\chi)$ (for example) will be represented by a discrete 
sine transform (DST), with an additional term linear in $\chi$.  Furthermore, 
the orbital source functions are not only periodic but have reflection 
symmetries across both periapsis ($\chi =0$) and apapsis ($\chi = \pi$).  
These symmetries narrow the form that the DCT can take to be either type I 
or II \cite{RaoYip90}.  We utilize the type I (referred to as DCT-I) 
algorithm with unitary normalization (making the DCT-I its own inverse).  

\begin{figure}
\includegraphics[scale=0.35]{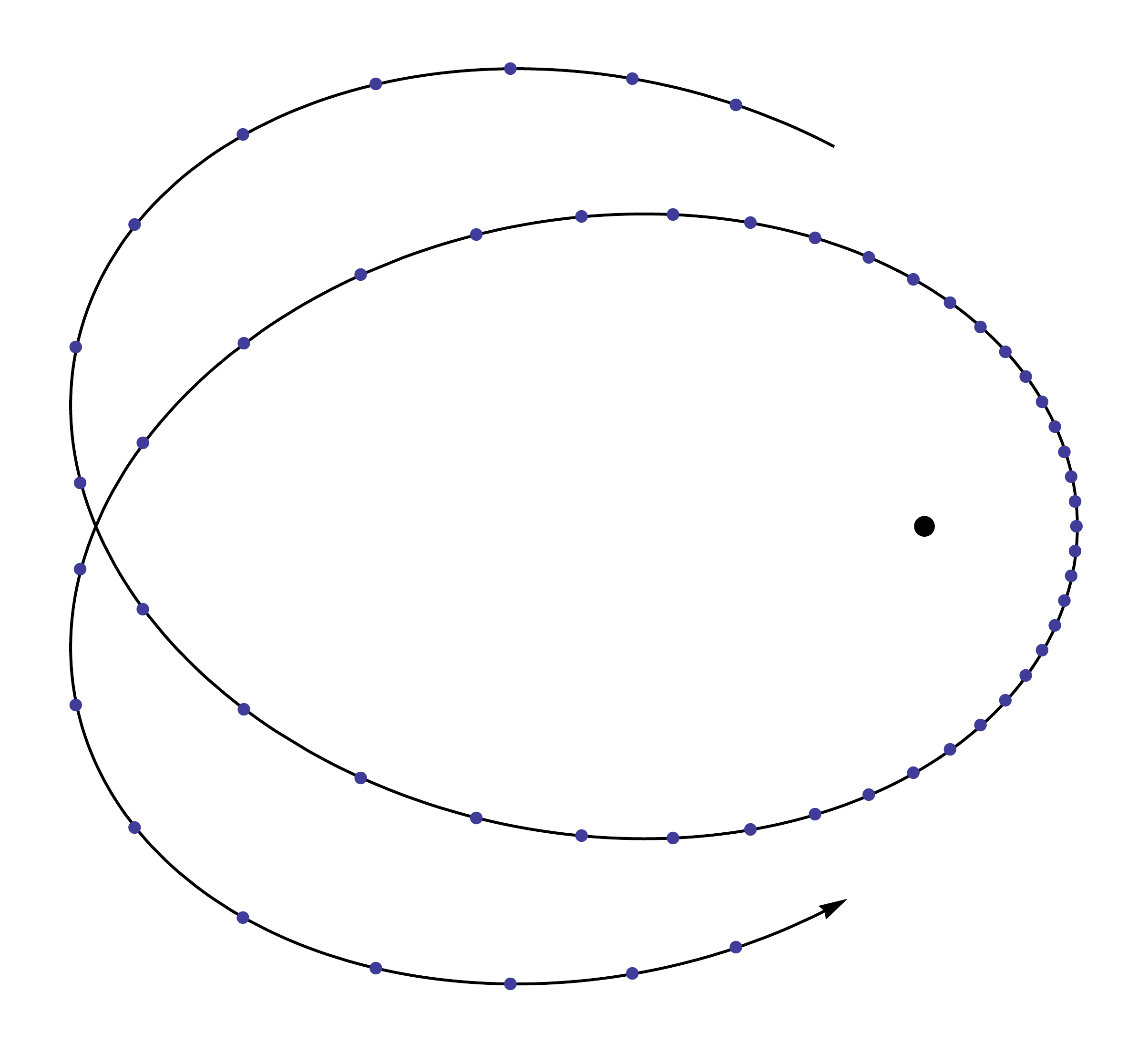}
\caption{
Equally spaced in $\chi$ sampling of $p=50$, $e=0.7$ orbit.  The complete 
orbit is split into $N = 42$ samples ($\Delta\chi = 0.1496$) and spectral 
integration requires only $\mathcal{N} = 22$ points between $\chi = 0$ and 
$\chi = \pi$ (inclusive) to achieve double precision accuracy.  The values 
of $dt_p/d\chi$ need only be calculated at the indicated points to provide 
double precision integration and interpolation anywhere on the orbit.
\label{fig:orbit50}} 
\end{figure}

In the general discussion above, we imagined dividing the entire orbit into 
$N$ intervals with $\Delta\chi = 2 \pi/N$.  For the DCT-I, this spacing is 
maintained and (assuming $N$ is even) the half orbit from $\chi = 0$ to 
$\chi = \pi$ is split into $N/2$ intervals.  The DCT-I utilizes 
$\mathcal{N} = N/2 + 1$ sample points by including the end points at both 
$\chi = 0$ and $\chi = \pi$.  In terms of the number of samples the domain 
is split into $\mathcal{N} - 1$ intervals.  The locations of the samples are
\begin{align}
\chi_k \equiv \frac{k \pi}{\mathcal{N}-1},	
\q \q 
k \in {0, 1, \ldots, \mathcal{N}-1}.
\end{align}
Then at the $\mathcal{N}$ points we denote the samples of the source function 
as $g_k = g(\chi_k)$.  The (real) Fourier coefficients are given by 
\begin{align}
\mathcal{G}_n &= \sqrt{\frac{2}{\mathcal{N}-1}} 
\bigg[ \frac{1}{2} g_0 + \frac{1}{2} (-1)^n g_{\mathcal{N}-1} \\ \notag 
& \hspace{24ex}
+ \sum_{k=1}^{\mathcal{N}-2} g_k \cos \left( n \chi_k \right) \bigg].
\end{align}
Like the more general DFT, this expression is exactly invertible and we can 
recover the original samples in the $\chi$-domain
\begin{align}
\label{eqn:gkDCT}
g_k &= \sqrt{\frac{2}{\mathcal{N}-1}} 
\bigg[ \frac{1}{2} \mathcal{G}_0 
+ \frac{1}{2} (-1)^k \mathcal{G}_{\mathcal{N}-1} \\ \notag 
& \hspace{24ex}
+ \sum_{n=1}^{\mathcal{N}-2} \mathcal{G}_n \cos \left( n \chi_k \right) \bigg].
\end{align}
We can then use the spectral amplitudes to provide a Fourier interpolation 
to arbitrary $\chi$
\begin{align}
\label{eqn:gContinuous}
g(\chi) &= \sqrt{\frac{2}{\mathcal{N}-1}} 
\bigg[ \frac{1}{2} \mathcal{G}_0 + \frac{1}{2} \mathcal{G}_{\mathcal{N}-1} 
\cos\left[(\mathcal{N}-1) \chi\right] \notag \\
& \hspace{22ex}
+ \sum_{n=1}^{\mathcal{N}-2} \mathcal{G}_n \cos \left( n \chi \right) \bigg].
\end{align}
Integrating Eqn.~\eqref{eqn:gContinuous} yields the sine expansion for the
time
\begin{align}
\label{eqn:tkDCT}
t_p(\chi) &= \sqrt{\frac{2}{\mathcal{N}-1}} 
\bigg[ \frac{1}{2} \mathcal{G}_0 \, \chi 
+ \frac{1}{2} \mathcal{G}_{\mathcal{N}-1} 
\frac{\sin \left[(\mathcal{N}-1) \chi\right]}{(\mathcal{N}-1)} \notag \\
& \hspace{21ex}
+ \sum_{n=1}^{\mathcal{N}-2} \frac{1}{n} \mathcal{G}_n 
\sin \left( n \chi \right) \bigg].
\end{align}

\begin{figure}
\includegraphics[scale=1.0]{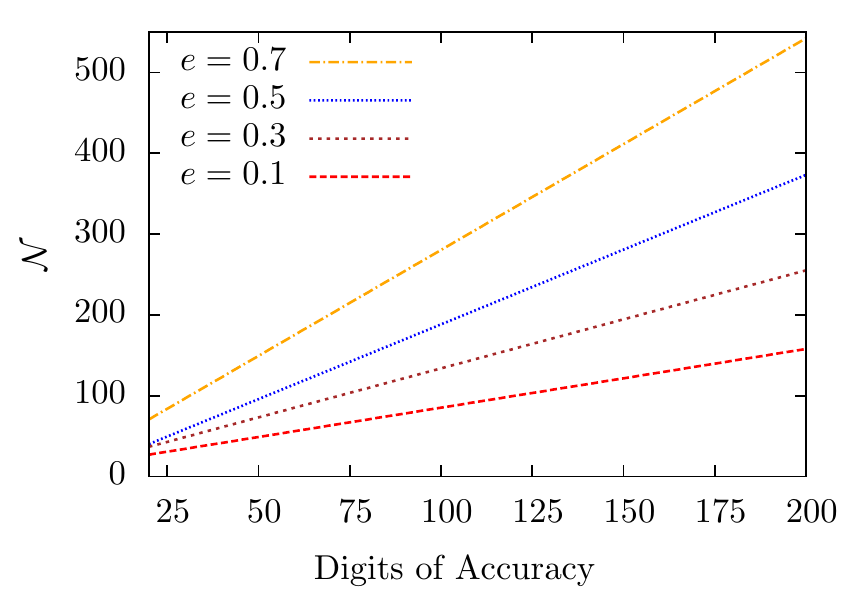}
\caption{
Number of sample points $\mathcal{N}$ between $\chi = 0$ and $\chi = \pi$ 
needed to represent $dt_p/d\chi = g(\chi)$ to a prescribed accuracy.  The 
ratio of magnitudes of the smallest to largest Fourier coefficients of 
$g(\chi)$ gives an estimate of the relative accuracy.  The linear scaling of
$\mathcal{N}$ versus digits of accuracy indicates geometric fall-off in the 
spectral components of $g(\chi)$.  Away from the separatrix this relation is 
largely independent of $p$.
\label{fig:samples}} 
\end{figure}

\begin{figure}
\includegraphics[scale=1.0]{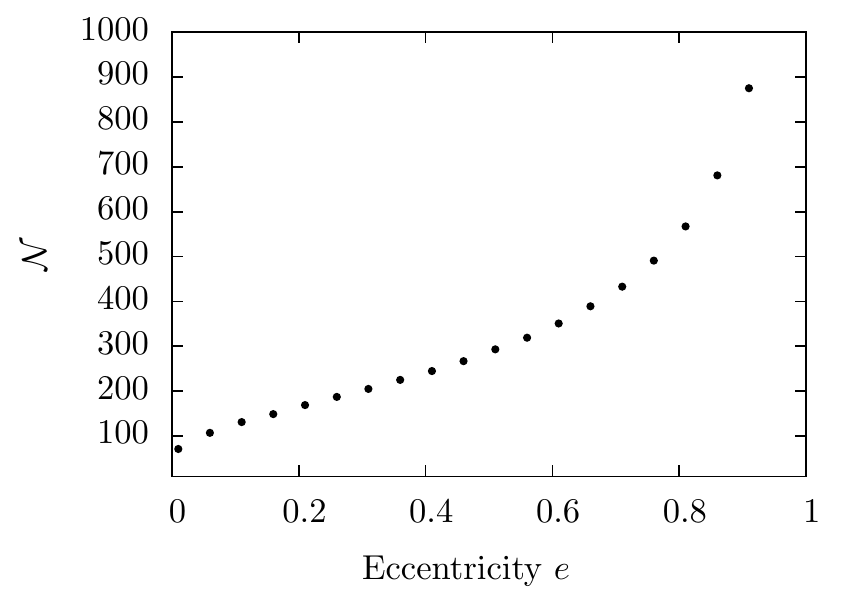}
\caption{
Number of sample points $\mathcal{N}$ between $\chi = 0$ and $\chi = \pi$ 
needed to represent $dt_p/d\chi = g(\chi)$ as a function of eccentricity $e$ 
at fixed accuracy of $150$ decimal places. Of course as the eccentricity
approaches unity and the radial period becomes infinite, so too does
the number of needed samples. Still, for any astrophysically relevant
orbit, we can sample the orbit to impressive accuracy with a modest number
of points.
\label{fig:samples_e}} 
\end{figure}

Having found $t_p (\chi)$ we can obtain the radial period $T_r$ from the 
leading Fourier amplitude $\mathcal{G}_0$
\begin{align}
T_r = \sqrt{\frac{2}{\mathcal{N}-1}} \pi \mathcal{G}_0 .
\end{align}
Then, with $T_r$ in hand, the fundamental frequencies can be computed
\begin{align}
\O_r = \frac{2 \pi}{T_r} ,
\q \q
\O_{\vp} = \frac{\vp_p(2 \pi)}{T_r} .
\end{align}

From a practical perspective, the DCT-I can be computed with 
$\mathcal{O} (\mathcal{N} \ln \mathcal{N})$ speed using either the 
FourierDCT function in \emph{Mathematica} or the FFTW routine in C coding.  
Fig.~\ref{fig:orbit50} provides a picture of how efficient this method is.  
For this orbit we need only $\mathcal{N} = 22$ samples to achieve double 
precision accuracy in the orbit integration.  In fact, all we need know are 
the source functions at the indicated points and we can interpolate to 
double precision accuracy anywhere in between.  From a practical standpoint, 
we guess a value of $\mathcal{N}$ and estimate the error by computing the 
ratio of the smallest to largest Fourier coefficients 
$|\mathcal{G}_{\mathcal{N}-1}/\mathcal{G}_0|$.  If that ratio fails to
meet our prescribed accuracy goal, we simply increase $\mathcal{N}$ and 
repeat the procedure.  Given that the DCT is so fast to compute, we are 
able to solve the orbit equations to hundreds of digits of accuracy 
within a few seconds.  Fig.~\ref{fig:samples} shows how remarkable and modest 
the scaling is in the number of required sample points $\mathcal{N}$ as a 
function of prescribed accuracy. Fig.~\ref{fig:samples_e} shows how
the number of needed sample points grows with increasing eccentricity
(given a fixed accuracy goal). Even at very high eccentricities, 
$e \approx 0.9$, the number of samples is quite reasonable.  Thus, with this 
approach the integration of the orbit becomes a trivial cost, even for 
astrophysically interesting eccentricities ($e \simeq 0.7$) and high 
accuracy (MST code) applications.

\section{Spectral source integration in the RWZ formalism}
\label{sec:method}

One of the principal goals of this paper is to describe our new means of 
applying spectral techniques (i.e., SSI) to integrate the source problem in 
black hole perturbation theory to high accuracy.  In this section we show 
the simplest application of SSI, namely solution of master equations in the 
Regge-Wheeler-Zerilli (RWZ) formalism for generic orbits.  Before detailing 
the SSI technique in Sec.~\ref{sec:SSI}, we first briefly review the now 
standard way \cite{HoppEvan10} of solving master equations using FD 
decomposition and the method of extended homogeneous solutions (EHS) 
\cite{BaraOriSago08}, and in the process set the notation. 

\subsection{The RWZ formalism and EHS method}
\label{sec:RWZEHS}

We begin with a RWZ master equation in the TD
\begin{align}
\label{eqn:TDmastereqn}
\left(-\frac{\pa^2}{\pa t^2} + \frac{\pa^2}{\pa r_*^2} - V_l(r)\right) 
\Psi_{lm}(t,r) = S_{lm}(t,r) ,
\end{align}
where $r_* = r + 2M \ln (r/2M - 1)$ is the usual tortoise coordinate.  Here 
$V_l (r)$ is either the Zerilli potential ($l+m$ even) or the Regge-Wheeler
potential ($l+m$ odd).  The source contains terms proportional to the Dirac 
delta function and its first derivative
\begin{align}
\begin{split}
\label{eqn:sourceTD}
& S_{lm}(t,r) = G_{lm}(t) \, \delta[r - r_p(t)] \\ 
& \hspace{25ex} + F_{lm}(t) \,\delta'[r - r_p(t)] .	
\end{split}
\end{align}
The time dependent functions $G_{lm}(t)$ and $F_{lm}(t)$ arise from tensor 
spherical harmonic decomposition \cite{HoppEvan10} of the stress-energy 
tensor of the point mass and enforcement of the delta function constraints 
$r \rightarrow r_p(t)$ and $\varphi \rightarrow \varphi_p(t)$.  Like the 
potential, their form depends upon parity.  For $l+m$ even we use the 
Zerilli-Moncrief source, and for $l+m$ odd we use the 
Cunningham-Price-Moncrief source (see \cite{HoppEvan10} for details).

As explained in Sec.~\ref{sec:orbit}, the eccentric motion of the source is
characterized by two fundamental frequencies, $\O_\varphi$ and $\O_r$.  As 
such, we can represent the master function and the source as Fourier series
\begin{align}
\label{eqn:psiSeries}
\Psi_{lm}(t,r) &= \sum_{n=-\infty}^\infty X_{lmn}(r) \, e^{-i \o t} , \\
S_{lm}(t,r) &= \sum_{n=-\infty}^\infty Z_{lmn}(r) \, e^{-i \o t} ,
\label{eqn:Slm}
\end{align}
with the mode frequencies being functions of both fundamentals 
\be
\label{eq:omega_mn}
\o = \o_{mn} \equiv m \O_\varphi + n \O_r, \q \q m,n \in \mathbb{Z}.
\ee
The series coefficients are formally found by integrating the TD master 
functions over one radial period
\begin{align}
X_{l mn}(r) &\equiv \frac{1}{T_r} \int_0^{T_r} dt \ \Psi_{l m}(t,r) 
\, e^{i \o t},
\\
Z_{l mn}(r) &\equiv \frac{1}{T_r} \int_0^{T_r} dt \ S_{l m}(t,r) 
\, e^{i \o t} .
\label{eqn:Zlmn}
\end{align}
The master equation then takes on the following FD form
\be
\label{eqn:FDmastereqn}
\left(\frac{d^2}{dr_*^2} +\o^2 -V_l(r)\right)
X_{lmn}(r) = Z_{lmn}(r).
\ee
(Throughout Sec.~\ref{sec:method} we do \emph{not} suppress any of the mode 
labels, though for all intents and purposes $l$ and $m$ can be regarded as 
fixed and arbitrary.)

An essential element in solving \eqref{eqn:TDmastereqn} is to obtain 
independent homogeneous solutions to \eqref{eqn:FDmastereqn}, either through 
numerical integration (after setting causal boundary conditions at 
$r_* \to \infty$ and $r_* \to -\infty$) or through use of analytic 
function (MST) expansions \cite{SasaTago03}.  We denote these unnormalized 
solutions as $\hat{X}^\pm_{lmn}(r)$, where
\begin{align}
\hat X_{lmn}^+ (r_* \to +\infty) & \sim e^{i \o r_*}, \\
\hat X_{lmn}^- (r_* \to -\infty) & \sim e^{-i \o r_*}.
\end{align}
A Green function is formed from these two linearly independent solutions 
and integrated over the source function $Z_{lmn}(r)$ to obtain the 
particular solution of \eqref{eqn:FDmastereqn}
\be
\label{eqn:FDInhomog}
X_{lmn} (r) = c^+_{lmn}(r) \, \hat{X}^+_{lmn}(r)
+ c^-_{lmn}(r) \, \hat{X}^-_{lmn}(r) ,
\ee
where the normalization functions in the source region are given by the 
integrals
\begin{align}
\label{eqn:cPM}
\begin{split}
c^+_{lmn}(r) &= \frac{1}{W_{lmn}} \, \int_{r_{\rm min}}^r   
\frac{dr'}{f(r')} \hat{X}^-_{lmn}(r') \, Z_{lmn}(r') , 
\\
c^-_{lmn}(r) &= \frac{1}{W_{lmn}} \, \int_r^{r_{\rm max}} 
\frac{dr'}{f(r')} \hat{X}^+_{lmn}(r') \, Z_{lmn}(r') .
\end{split}
\end{align}
Here $W_{lmn}$ is the Wronskian 
\be
W_{lmn} = f(r) \left( \hat{X}^-_{lmn} \frac{d \hat{X}^+_{lmn}}{dr}
- \hat{X}^+_{lmn} \frac{d \hat{X}^-_{lmn}}{dr} \right) .
\ee

While the expression in Eqn.~\eqref{eqn:FDInhomog} is indeed a solution to 
Eqn.~\eqref{eqn:FDmastereqn}, it is not ideal.  The singular nature of the 
TD source \eqref{eqn:sourceTD} results in Gibbs behavior in the Fourier 
synthesis \eqref{eqn:psiSeries} of $\Psi_{lm}$ at and near the particle 
location, leading to slow algebraic convergence.  Exponential convergence 
can be restored by using the method of EHS, originally developed by Barack, 
Ori and Sago \cite{BaraOriSago08}.  

The first step in EHS is to extend the limits of integration in 
\eqref{eqn:cPM} to include the full source region and obtain the 
normalization coefficients
\be
\label{eqn:normC}
C_{lmn}^{\pm} 
= \frac{1}{W_{lmn}} \int_{r_{\rm min}}^{r_{\rm max}} dr
 \ \frac{\hat X^{\mp}_{lmn} (r) Z_{lmn} (r)}{ f(r)} .
\ee
These complex constants are in turn used to normalize the individual mode 
functions
\be
\label{eq:FD_EHS}
X^\pm_{lmn} (r) 
= C^{\pm}_{lmn} \hat X_{lmn}^\pm (r) , 
\ee
producing the FD EHS of Eqn.~\eqref{eqn:FDmastereqn}.  Collectively, these 
normalized modes encode all the information about the source motion 
and are used to then define the TD EHS
\be
\label{eq:TD_EHS}
\Psi^\pm_{lm} (t,r) 
\equiv \sum_n X^\pm_{lmn} (r) \, e^{-i \o t} .
\ee
As the FD EHS are each $C^\infty$, these Fourier sums converge exponentially 
for all $r > 2 M$.  The sums are formally infinite in number, but in practice 
they are truncated once a specified accuracy is reached.  The desired 
particular TD solution to Eqn.~\eqref{eqn:TDmastereqn} is then obtained by 
joining the outer and inner TD EHS:
\be
\Psi_{lm}(t,r) = \Psi^{+}_{lm} \theta \left[ r - r_p(t) \right] + 
\Psi^{-}_{lm} \theta  \left[ r_p(t) - r \right] .
\ee
This weak solution can be computed everywhere, including the particle 
location, and it allows the metric and local gravitational self-force to be 
accurately determined \cite{HoppEvan10}.

There remains the practical issue of computing the $C_{lmn}^{\pm}$.  For 
the RWZ problem, the source $Z_{lmn}(r)$ in Eqn.~\eqref{eqn:normC} is 
poorly behaved at the turning points because of the presence of the 
$\delta'$ term in \eqref{eqn:sourceTD} \cite{HoppEvan10}.  It was shown in 
that paper that the problem could be circumvented by reversing the order 
of integration (see related examples in \cite{BaraOriSago08,WarbBara11}).  
To see this, substitute the Fourier transform integral for $Z_{lmn}(r)$ into 
\eqref{eqn:normC}
\begin{align}
\label{eq:Cpm2}
C_{lmn}^\pm  = &\frac{1}{W_{lmn} T_r} 
\int_{r_{\rm min}}^{r_{\rm max}} dr 
\ \frac{\hat X^\mp_{lmn} (r)}{f(r)} \hspace{8ex} \\ \notag
&\hspace{18ex} \times \int_0^{T_r} dt \ S_{lm}(t,r) e^{i \o t} . 
\end{align}
Then substitute for the TD source $S_{lm}(t,r)$ its singular form 
\eqref{eqn:sourceTD}, exchange the order of integration, and integrate in 
$r$ over the delta function terms.  What remains of the calculation of 
$C_{lmn}^{\pm}$ is an integral over time
\begin{align}
\label{eqn:EHSC}	
C_{lmn}^\pm  =  &\frac{1}{W_{lmn} T_r} \int_0^{T_r}
\Bigg[ 
 \frac{1}{f_{p}} \hat X^\mp_{lmn}
 G_{lm} \hspace{5ex} \\ \notag
&\hspace{5ex} 
+ \l \frac{2M}{r_{p}^2 f_{p}^{2}} \hat X^\mp_{lmn}
 - \frac{1}{f_{p}} 
 \frac{d \hat X^\mp_{lmn}}{dr} \r F_{lm}
 \Bigg]  e^{i \o t}  \, dt .
\end{align}
The integrand is composed of obvious functions of time, such as 
$G_{lm}(t)$ and $F_{lm}(t)$.  However, all of the other terms inside the 
square braces are now also functions of time, since the delta function maps 
$r \rightarrow r_p(t)$ [e.g., $f_p \equiv f(r_p(t))$, 
$\hat X^\mp_{lmn}(r) \rightarrow \hat X^\mp_{lmn}(r_p(t))$]. 

In summary, the RWZ BHP problem is solved by computing, for a sufficient 
range of $l$, $m$, and $n$, the inner and outer mode functions 
$\hat X_{lmn}^\pm (r)$ (by ODE integration or analytic function expansion) 
and computing the integrals \eqref{eqn:EHSC} for the normalization 
coefficients $C_{lmn}^{\pm}$ (using either IVP ODE integration 
\cite{HoppEvan10} or a numerical quadrature routine \cite{WarbBara11}).  

\subsection{SSI for the normalization coefficients}
\label{sec:SSI}

SSI is a new modification in the way the normalization coefficients 
$C_{lmn}^\pm$ are calculated.  The key first step in developing SSI was 
actually the reversal in the order of integration described immediately 
above.  The second essential step involves recognizing the periodic nature 
of the integrand in \eqref{eqn:EHSC}.  The functions $F_{lm}(t)$ and 
$G_{lm}(t)$, which contribute to the source $S_{lm}$, have complex time 
dependence because of the biperiodic motion and (typically) incommensurate 
frequencies $\Omega_r$ and $\Omega_\varphi$.  The motion in $\varphi$ can 
be split into
\be
\varphi_p(t) = \Omega_{\varphi} t + \Delta\varphi(t) ,
\ee
where the mean azimuthal advance is modulated by $\Delta\varphi(t)$, which 
is periodic in the radial motion.  This $\varphi_p(t)$ enters source terms 
only through the spherical harmonic factor $e^{-im \vp_p(t)}$, which factors
into: $e^{-im\Omega_{\vp} t} \, e^{-im \Delta\vp(t)}$.  It is the mean 
azimuthal phase advance, at angular rate $\O_\vp$, that makes source terms 
biperiodic.  We can, however, define functions $\bar{G}_{lm}$ and 
$\bar{F}_{lm}$ via
\begin{align}
\begin{split}
\label{eqn:GFBar}
\bar{G}_{lm}(t) &\equiv G_{lm}(t) \, e^{im\Omega_{\vp} t}, \\
\bar{F}_{lm}(t) &\equiv F_{lm}(t) \, e^{im\Omega_{\vp} t},
\end{split}
\end{align}
that are strictly $T_r$-periodic.  Returning to Eqn.~\eqref{eqn:EHSC}, we see 
that the factor, $e^{-im\Omega_{\vp} t}$, responsible for biperiodicity, 
cancels with a corresponding factor from the Fourier transform kernel.  
We can replace the integral with
\begin{align}
C_{lmn}^\pm  
&= \frac{1}{W_{lmn} T_r} \int_0^{T_r}
\bar{E}_{lmn}^{\pm}(t) \ e^{i n\O_r t} \ dt.
 \label{eqn:CfromEbarT}
\end{align}
where $\bar{E}^{\pm}_{lmn}(t)$ are strictly $T_r$-periodic functions 
\begin{align}
\label{eqn:eBarDef}
\bar{E}^{\pm}_{lmn}(t)
&\equiv \frac{1}{f_{p}} \hat X^\mp_{lmn}
\bar{G}_{lm} \hspace{5ex} \\ \notag
&\hspace{4ex} 
+ \l \frac{2M}{r_{p}^2 f_{p}^{2}} \hat X^\mp_{lmn}
 - \frac{1}{f_{p}} 
 \frac{d \hat X^\mp_{lmn}}{dr} \r \bar{F}_{lm} .
\end{align}

The third, and most important, step toward SSI harks back to our earlier 
discussion in Sec.~\ref{sec:orbitSpec} of spectrally integrating the orbit 
equations.  There we showed that due to the $C^\infty$ smoothness of (for 
example) $dt_p/d\chi = g(\chi)$ we could replace $g(\chi)$ with an 
equally-spaced sampling $g_k = g(k \Delta\chi)$ of modest total number of 
samples $N$ and achieve high-accuracy interpolation and integration.  For 
source integration, the equivalent step (to be justified momentarily) is to 
replace \eqref{eqn:CfromEbarT} with
\begin{align}
\label{eqn:Cseries}
C_{lmn}^\pm  
&= \frac{1}{N W_{lmn}} \sum_{k=0}^{N-1} \bar{E}^{\pm}_{lmn}(t_k) 
\ e^{i n \O_r t_k } ,
\end{align}
where the time samples are $t_k = k T_r/N$, with $k = 0,\ldots ,N-1$.  
\emph{This remarkably simple sum is the heart of SSI.}  By replacing the 
integral in \eqref{eqn:CfromEbarT} with the sum in \eqref{eqn:Cseries}, we 
avoid ODE integration and the calculation of the normalization coefficients 
is vastly sped up, opening the door to much higher accuracy 
applications \cite{ForsEvanHopp15}.  

\emph{What makes SSI work?}  Before we examine how well SSI performs, we first 
justify \eqref{eqn:Cseries} as an appropriate replacement 
for \eqref{eqn:CfromEbarT}.  The argument starts by noting the expected 
smoothness of the functions $\bar{E}_{lmn}^{\pm}(t)$ that enter 
\eqref{eqn:CfromEbarT}.  The contributing elements $\bar{F}_{lm}(t)$ and 
$\bar{G}_{lm}(t)$ are smooth $C^\infty$ functions of the orbital motion.  
Similarly, the modes $\hat{X}_{lmn}^{\mp}(r)$ are smooth functions of $r$, 
and hence become smooth functions of time under the replacement 
$r \rightarrow r_p(t)$.  Thus, for every $lmn$, the integrand in 
\eqref{eqn:CfromEbarT} is smooth and periodic.  These properties suggest, 
just as they did in Sec.~\ref{sec:SpecInt}, use of FS expansion.  Indeed, 
the integral in Eqn.~\eqref{eqn:CfromEbarT} looks like, under a cursory 
glance, the calculation of a set of FS coefficients.  However, it is clear 
that $C_{lmn}^\pm$ is not a spectrum of coefficients (in $n$) derived from 
a single function of time, but is instead calculated from a whole set 
(in $n$) of TD functions $\bar{E}_{lmn}^{\pm}(t)$.

Nevertheless, the Fourier series can be put to investigative use and we 
introduce one for each $\bar{E}_{lmn}^{\pm}(t)$:
\be
\label{eqn:EFS}
\bar{E}_{lmn}^{\pm}(t) = \sum_{n'=-\infty}^\infty 
\ti{\mathcal{E}}_{lmnn'}^{\pm} \ e^{-i n' \Omega_r t} , 
\ee
with the coefficients given by
\be
\label{eqn:EFScoeff}
\ti{\mathcal{E}}_{lmnn'}^{\pm} = 
\frac{1}{T_r} \int_0^{T_r} dt \ \bar{E}_{lmn}^{\pm}(t) 
\, e^{i n' \Omega_r t} .
\ee
If \eqref{eqn:EFS} is substituted in \eqref{eqn:CfromEbarT}, and sum and 
integral are exchanged, we find that the normalization coefficients
\be
\label{eqn:Enn}
C_{lmn}^\pm = \frac{1}{W_{lmn}} \ \ti{\mathcal{E}}_{lmnn}^{\pm} \quad ,
\ee
are proportional to the diagonal elements ($n = n'$) of the superset (over 
$n$ and $n'$) of FS coefficients $\ti{\mathcal{E}}_{lmnn'}^{\pm}$.  The 
result is understandable: the integral in \eqref{eqn:CfromEbarT} simply 
picks out the $n$th harmonic in the $n$th function $\bar{E}_{lmn}^{\pm}(t)$.

To complete the argument, we may assume (and numerically verify) that the 
smoothness of a source function $\bar{E}_{lmn}^{\pm}(t)$ implies a rapidly 
falling (likely geometric) spectrum for $\ti{\mathcal{E}}_{lmnn'}^{\pm}$ as 
$n' \rightarrow \pm \infty$.  As we argued in Sec.~\ref{sec:SpecInt}, 
for any given accuracy goal, this implies the spectrum can be truncated at 
some sufficiently negative and positive values of $n'$.  Truncation, in 
turn, means that we have replaced the original source function with a
bandlimited approximation.  Bandlimiting then argues for 
replacing the source function (yet again), this time with a set of discrete, 
equally-spaced samples $\bar{E}_{lmn}^{\pm}(t_k)$.  Because the source 
function is periodic, the discrete sampling is finite in number (say $N$).  
We can then use the DFT to relate the discrete sampling representation of 
the source to a discrete, finite spectrum (and vice versa)
\be
\label{eqn:DFTback}
\bar{E}_{lmn}^{\pm}(t_k) = \sum_{n' = 0}^{N-1} 
\mathcal{E}_{lmnn'}^{\pm} \ e^{-i n' \Omega_r t_k} , 
\ee
\be
\label{eqn:DFTforward}
\mathcal{E}_{lmnn'}^{\pm} = \frac{1}{N} \sum_{k = 0}^{N-1} 
\bar{E}_{lmn}^{\pm}(t_k) \ e^{i n' \Omega_r t_k} .
\ee
The DFT spectrum $\mathcal{E}_{lmnn'}^{\pm}$ is distinct from the FS 
spectrum $\ti{\mathcal{E}}_{lmnn'}^{\pm}$, and the former will display 
periodicity in the FD, 
$\mathcal{E}_{lmn,n'+jN}^{\pm} = \mathcal{E}_{lmnn'}^{\pm}$.  
However, for sufficiently large $N$ and between the negative and positive 
Nyquist frequencies, the two spectra can be made nearly indistinguishable.  
If we then set $n' = n$, replace $\ti{\mathcal{E}}_{lmnn}^{\pm}$ in 
\eqref{eqn:Enn} with the DFT spectral component $\mathcal{E}_{lmnn}^{\pm}$, 
and substitute into the same equation the DFT relation 
\eqref{eqn:DFTforward}, we have derived our SSI formula 
Eqn.~\eqref{eqn:Cseries}.

We can provide a summary of this discussion, and the derivation, through 
a sequence of replacements:
\begin{align}
C_{lmn}^\pm &= \frac{1}{W_{lmn} T_r}
\int_0^{T_r} dt \ \bar{E}_{lmn}^{\pm}(t) \ e^{i n\O_r t} \\ \notag
&= \frac{1}{W_{lmn} T_r}
\int_0^{T_r} dt \ e^{i n\O_r t} \, \sum_{n'=-\infty}^\infty 
\ti{\mathcal{E}}_{lmnn'}^{\pm} \ e^{-i n' \Omega_r t} \\ \notag
&\simeq \frac{1}{W_{lmn} T_r}
\int_0^{T_r} dt \ e^{i n\O_r t} \,
\sum_{n'= n_{\rm min}^\prime}^{n_{\rm max}^\prime}
\ti{\mathcal{E}}_{lmnn'}^{\pm} \ e^{-i n' \Omega_r t} \\ \notag
&\simeq \frac{1}{W_{lmn} T_r}
\int_0^{T_r} dt \ e^{i n\O_r t} \,
\sum_{n'= n_{\rm min}^\prime}^{n_{\rm max}^\prime}
\mathcal{E}_{lmnn'}^{\pm} \ e^{-i n' \Omega_r t} \\ \notag
&= \frac{1}{W_{lmn} T_r} 
\int_0^{T_r} dt \ e^{i n\O_r t} \,
\sum_{n'= n_{\rm min}^\prime}^{n_{\rm max}^\prime}
\, e^{-i n' \Omega_r t} \\ \notag 
&\qquad\qquad\qquad\qquad \times \frac{1}{N} \sum_{k = 0}^{N-1} 
\bar{E}_{lmn}^{\pm}(t_k) \ e^{i n' \Omega_r t_k} \\ \notag
&=\frac{1}{N W_{lmn}} 
\, \sum_{n'= n_{\rm min}^\prime}^{n_{\rm max}^\prime}
\sum_{k = 0}^{N-1} \bar{E}_{lmn}^{\pm}(t_k) \ e^{i n' \Omega_r t_k} 
\ \delta_{nn'} \\ \notag
&= \frac{1}{N W_{lmn}} \, \sum_{k = 0}^{N-1} \bar{E}_{lmn}^{\pm}(t_k) 
\ e^{i n \Omega_r t_k} .
\end{align}
The two approximate (but typically spectrally accurate) steps are indicated.

\begin{figure}
\center
\includegraphics[scale=1]{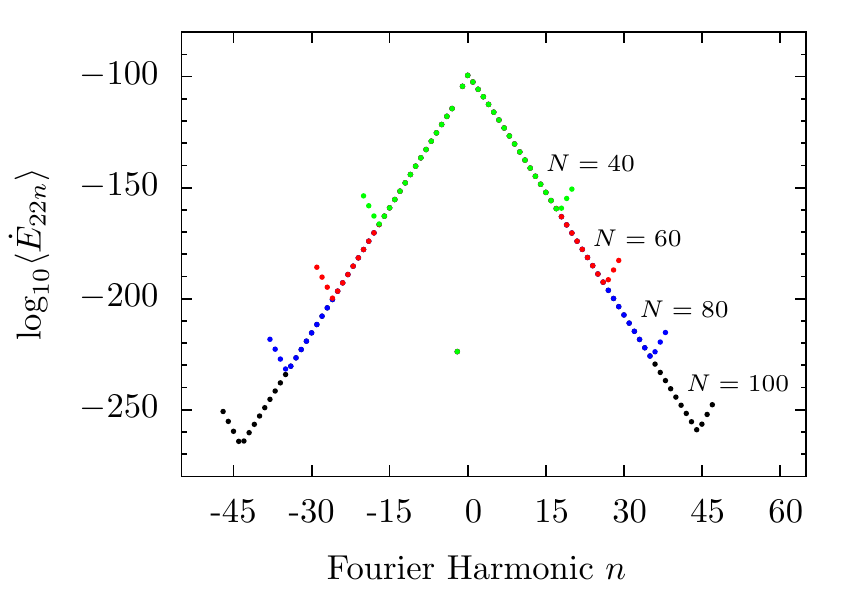}
\caption{
Aliasing effect from oversampling SSI in the FD.  Shown here are energy 
flux data from an orbit with $p=10^{20}$ and $e=0.01$ for modes with $l=2$, 
$m=2$.  The energy flux from successive $n$ modes falls off exponentially 
when computed away from the peak harmonic.  Note that one harmonic 
($n=-m=-2$) is nearly static, which decreases its flux by more than $100$ 
orders of magnitude.  As higher positive and negative $n$ are computed, the 
fluxes reach Nyquist-like notches and oversampling in $n$ beyond those 
points leads to increases in flux similar to aliasing.  The locations of 
the minima scale with but are not equal to $\pm N/2$.
}
\label{fig:alias}
\end{figure}

\emph{What is involved in practical use of SSI?}  Another way of asking this 
question is: if we make $N$ discrete samples of each source function and 
sum them in \eqref{eqn:Cseries}, for which and how many $n$'s should we 
compute $C_{lmn}^{\pm}$?  We do not presently have an exact answer, but we 
have an effective, practical procedure.  To see the issue, consider 
Fig.~\ref{fig:alias}.  There we show gravitational wave 
energy fluxes per harmonic $n$ at $r = \infty$ for the $l=2$, $m=2$ mode 
(essentially proportional to $|C_{22n}^{+}|^2$).  We might expect, for a 
given $N$, to begin near $n=0$ and see a spectrum that descends on either side 
until hitting a Nyquist point (at about $n = \pm N/2$).  That is roughly, 
but not exactly, what is observed.  The problem is that $C_{lmn}^{\pm}$ is 
not, as a function of $n$, a DFT spectrum.  If we consider \eqref{eqn:Enn}, 
clearly the Wronskian $W_{lmn}$ should not be expected to display a 
periodicity in $n$.  Even the DFT spectra, while having the periodicity in 
$n'$, $\mathcal{E}_{lmn,n'+jN}^{\pm} = \mathcal{E}_{lmnn'}^{\pm}$, will not 
have a periodicity in the diagonal elements $\mathcal{E}_{lmnn}^{\pm}$ as 
a function of $n$.  Nevertheless, if we sample $C_{lmn}^{\pm}$ in $n$ for 
$|n| \gtrsim N/2$ we observe a succession of Nyquist-like notches and peaks, 
similar to aliasing in the DFT but without exact periodicity.  From a practical 
standpoint, we compute and use the spectrum in $n$ down to the first 
Nyquist-like notch on each side and calculate no further.  The code marches 
forward on each side, finds the minima, and discards contributions beyond 
those points.

\begin{figure}
\center
\includegraphics[scale=1]{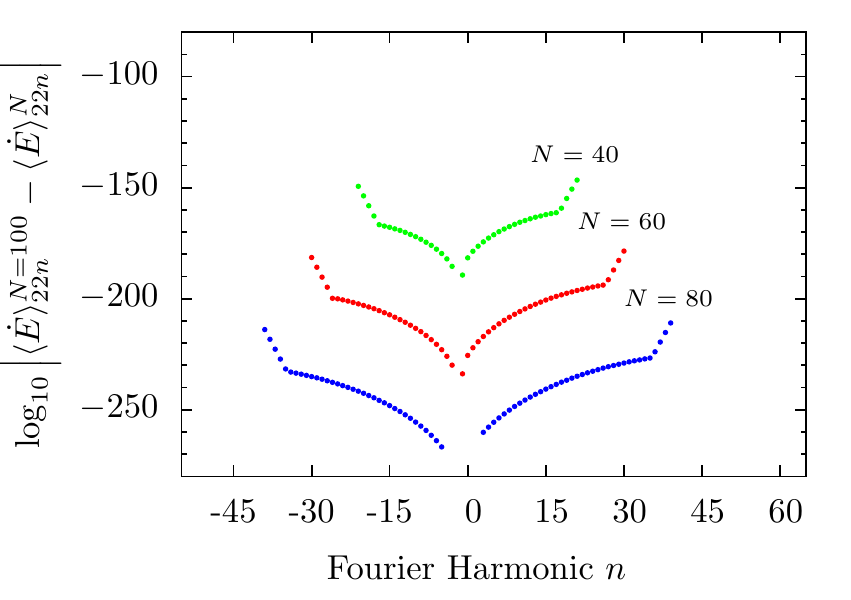}
\caption{
SSI at high accuracy.  Absolute differences (errors) are shown in 
self-convergence tests.  The same data found in Fig.~\ref{fig:alias} are
used to compute differences (per harmonic) in the fluxes between the three 
lower resolutions and the highest ($N = 100$) resolution.  For each of the 
three lower resolutions ($N = 40,60,80$) the errors are well bounded by the 
accuracy criteria set by errors at the Nyquist notches.
}
\label{fig:alias2}
\end{figure}

Fig.~\ref{fig:alias} shows this aliasing phenomenon.  There we deliberately 
compute and display energy fluxes for a few harmonics beyond the first 
Nyquist notch on each side of the central maximum.  We show the same fluxes 
computed with four different spectral resolutions.  The exponential fall in 
the spectrum is evident.  These calculations were made possible not only by 
use of SSI but also \emph{Mathematica's} arbitrary precision arithmetic.  
As $N$ becomes larger, we approach the FS, or continuum, limit.  
It is clear from the vantage point of high resolution that the best thing 
to do at lower resolution is halt the mode calculations at the Nyquist 
notches.  This assumption is borne out by considering Fig.~\ref{fig:alias2}.  
This figure displays the differences in fluxes between those computed at 
resolutions of $N = 40, 60, 80$ and those found with $N = 100$.  The error 
in the discrete representation is well bounded in the region between the 
first Nyquist points by the maximum error at one of the notches.

Operating at high accuracy (e.g., 200 decimal places), the MST code makes a
prediction of how large $N$ needs to be in order that the Nyquist notches lie 
(just) below the specified error level.  We have observed that adequate 
sampling for SSI is always sufficient for comparably accurate orbit 
integration.  The prediction for $N$ is tested and if mode fluxes do not 
reach the error level at the Nyquist notches, then a new value of $N$ is 
chosen and the calculation is repeated.  In the MST code, the now vastly 
reduced number of function evaluations can still be expensive if $N$ is set 
too generously.  It is important to note that the key formula for SSI, 
Eqn.~\eqref{eqn:Cseries} (or more properly \eqref{eqn:CseriesChi}--see below), 
is an $\mathcal{O}(N^2)$ procedure, and so it is
essential to find a near minimum 
value of $N$ for a given accuracy goal.  

\emph{How well does SSI work?}  We have demonstrated numerically in 
Figs.~\ref{fig:alias} and \ref{fig:alias2} the presence of exponential 
convergence.  It is not that the gravitational wave fluxes fall geometrically
(a known result), but that the gap between resolutions (i.e., error in 
substituting the DFT for the Fourier series) falls exponentially with 
increases in $N$.  We can, however, go a step further and make the rate of 
exponential convergence even faster by introducing one final modification.  

SSI is exponentially convergent because the periodic functions, 
$\bar{E}^{\pm}_{lmn}(t)$, being sampled are $C^{\infty}$.  However, there 
is no requirement that the periodic motion be described by $t$.  Any 
$C^\infty$ reparametrization $t \rightarrow \lambda(t)$ should be expected to 
also give rapidly convergent sums.  This is true, for example, in switching 
from $t$ to the relativistic anomaly $\chi$.  We have found empirically, 
though, that use of $\chi$ as the curve parameter substantially improves
the rate of exponential convergence.

\begin{figure}
\center
\includegraphics[scale=1]{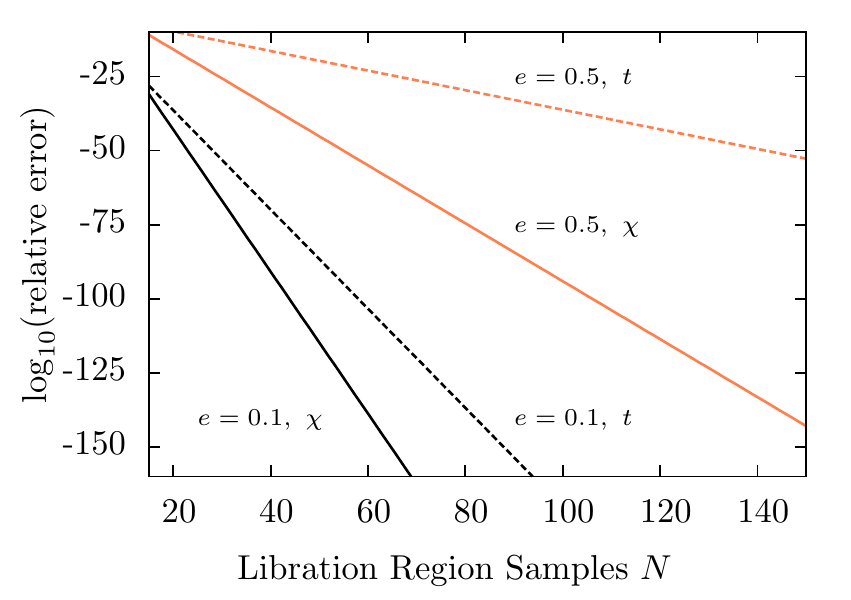}
\caption{
Comparison between SSI with $\chi$ sampling and $t$ sampling as a function 
of $N$ and eccentricity.  The two sampling schemes are tested by examining 
convergence of the $l=2$, $m=2$, $n=0$ energy flux using the MST code.  All 
orbits have $p = 10^{3}$.}
\label{fig:fluxMST}
\end{figure}

To effect this change we rewrite Eqn.~\eqref{eqn:CfromEbarT} with $\chi$ 
as the independent variable.  Then the periodic motion is divided into 
equally spaced steps $\Delta\chi = 2\pi/N$, the integrand is discretely 
sampled, and the integral is replaced with the sum 
\begin{align}
\label{eqn:CseriesChi}
C^{\pm}_{lmn} &= \frac{\O_r}{N W_{lmn}} \sum_k 
\frac{dt_p}{d\chi} \
\bar{E}^{\pm}_{lmn}[t(\chi_k)]
\, e^{i n\Omega_r t(\chi_k)} .
\end{align}
As Fig.~\ref{fig:fluxMST} demonstrates, substantially fewer $\chi$-samples 
are required than $t$-samples for SSI to reach a prescribed error level.  
This is especially true of high eccentricity orbits.  As a practical matter, 
it is also easier to find $\vp_p$ and $r_p$ evenly sampled in $\chi$ than 
in $t$.  Finally, we comment that it was merely a hunch (though one informed 
by experience with the problem) that $\chi$ might provide a better measure 
and more rapid convergence.  It is an open question whether there is another 
parametrization of the orbit that yields even faster convergence.

\begin{figure*}[ht]
\center
\includegraphics[width=\textwidth,scale=1.15]{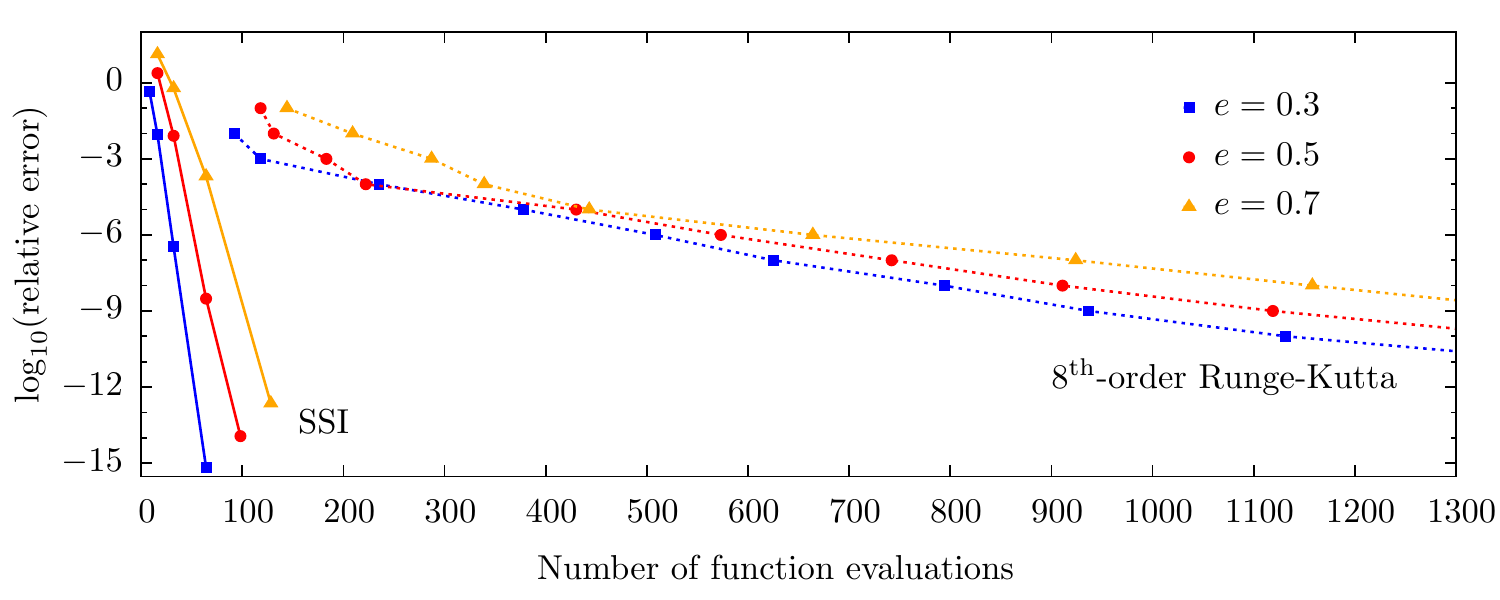}
\caption{
Efficiency of spectral source integration in comparison to ODE integration 
in a RWZ application at double precision.  The RWZ normalization constant 
$C^+_{220}$ is computed for various eccentricities in orbits with $p=10$.  The 
ODE integration uses the Runge-Kutta-Prince-Dormand 7(8)~\cite{PrinDorm81} 
routine rk8pd of the GNU Scientific Library (GSL)~\cite{GSL}. }
\label{fig:funceval}
\end{figure*}

The numerical SSI results shown so far have involved the high accuracy MST 
code.  But SSI also aids in double precision C coded calculations.  Its
benefit is shown clearly in Fig.~\ref{fig:funceval}, where we mark the 
accuracy reached in computing a normalization coefficient ($C_{220}^{+}$) 
as a function of the number of source term evaluations, which serves as a 
proxy for computational load.  We compare SSI to an IVP ODE integration using 
an 8th order Runge-Kutta routine.  SSI has exponentially converging accuracy 
with increases in function calls (i.e., increases in $N$).  In contrast, 
the Runge-Kutta routine, with its algebraic convergence, struggles to reach 
high accuracies.  

\emph{Do we really need SSI?}  The answer depends upon the application.  
At double precision the answer is clearly no, but SSI is likely much more 
efficient (and hence faster).  The real critical requirement for SSI comes in 
high accuracy eccentric orbit calculations.  Consider Fig.~\ref{fig:funceval} 
again and the scaling of ODE integration.  At an accuracy goal of $200$ 
digits, even an efficient algorithm like 8th order Runge-Kutta would take of 
order $10^{22}$ steps to integrate through an eccentric orbit source 
region!  Without SSI or a comparable spectral method, these calculations 
simply cannot be done.

\subsection{SSI and the midpoint and trapezoidal rules}
\label{sec:SSItrap}

We developed SSI with the convergence of Fourier series and concepts in 
signal processing (e.g., sampling theorem, use of the DFT/FFT, etc) firmly 
in mind.  The application 
of SSI to orbit integration does in fact simply use the DCT, a special case 
of the DFT.  In source integration, even though the key formula, 
\eqref{eqn:CseriesChi} or \eqref{eqn:Cseries}, is not a DFT, we used the 
DFT to provide an understanding of the rapid convergence of the sum.  The 
essential point was to see that rapid convergence in the FD with a modest 
number $N$ of spectral elements could translate into representing the 
behavior in the TD with equally modest sampling.  If the representation has 
sufficient accuracy, then interpolation and integration can be made 
accurately as well.

Yet, if we step back and examine the sums [\eqref{eqn:CseriesChi} or 
\eqref{eqn:Cseries}] that we use in SSI, a curious fact jumps out: they 
appear to be nothing more than simple Riemann sums.  Given the sampling, 
their form appears to be a use of the left rectangle rule.  However, with 
the inherent periodicity in $\chi$, the left rectangle rule is equivalent 
to the trapezoidal rule and, with a half interval shift in the equal-sized 
$\chi$ bins, it is also just the midpoint rule.  But these are just the 
lowest-order approximations for an integral, with error bounds, 
$\mathcal{O}(1/N^2)$, that are algebraic in the number of divisions of the 
interval!  How can their use be giving a vastly faster rate of convergence?  
The answer lies in the periodicity and smoothness of the summands.  After 
developing the method we came upon a paper \cite{Weid02} that discusses this 
surprising behavior in other contexts and nicely provides a set of example 
calculations.  A more recent and exhaustive discussion is found in 
\cite{TrefWeid14}.  In black hole perturbation work, Fujita, Hikida, and 
Tagoshi \cite{FujiHikiTago09} made use of the trapezoidal rule for source 
integration, but did not explicitly note or demonstrate the exponential 
convergence or push the results beyond double precision.
The remarkably rapid 
convergence of the trapezoidal rule in special cases is apparently well 
known in certain numerical analysis circles, though it also appears to be 
something that is constantly being rediscovered ever since Poisson's 
original finding in 1827 \cite{Pois1827}. 

Since we have shifted the viewpoint momentarily to thinking about Riemann 
sums and quadrature formulae, what about higher-order methods like Simpson's 
rule?  Since Simpson's rule generally has a stronger error bound (implying
presumably faster convergence) than the trapezoidal rule, might its use in 
SSI allow us to converge even faster?  Alas, the answer is no, as a quick 
test demonstrated.  A discussion and example can be found in \cite{Weid02}.  

\section{Spectral source integration in Lorenz gauge}
\label{sec:methodSystem}

After developing SSI for use in the RWZ formalism, we turned our attention
to Lorenz gauge and were able to successfully apply the method to coupled 
systems of equations.  Lorenz gauge breaks down into several systems of 
different orders that depend on (1) parity, (2) mode order (either low 
$l=0,1$ or high $l\ge 2$), and (3) the special static ($\omega = 0$) case.  
See \cite{AkcaWarbBara13} and \cite{OsbuETC14} for details.  Here we simply
demonstrate the principles of incorporating SSI by focusing only on the 
odd-parity equations.  

\subsection{Odd-parity Lorenz gauge and EHS}

In Lorenz gauge, odd-parity perturbations can be described by the two 
amplitudes, $h_t^{lm}$ and $h_r^{lm}$, with the third, $h_2^{lm}$, 
obtainable from the gauge condition.  The reduced-order coupled system in 
the TD is
\begin{align}
\label{eqn:EqnsConstrOddht}
& \Box h^{lm}_t
- \frac{2M f}{r^2} \pa_r h^{lm}_t
+ \frac{2Mf}{r^2} \pa_t h^{lm}_r  \\ \notag
& \hspace{12ex}
- \frac{2 f^2 + (l+2)(l-1) f}{r^2} h^{lm}_t 
 = f^2 P_{lm}^t, 
\\
\label{eqn:EqnsConstrOddhr}
& \Box h^{lm}_r 
+ \frac{2(r-M)f}{r^2} \pa_r h^{lm}_r 
- \frac{2(r-3M)}{r^2 f} \pa_t h^{lm}_t   \\ \notag
& \hspace{12ex}
- \frac{2 f^2 + (l+2)(l-1) f}{r^2}  h^{lm}_r 
= - P_{lm}^r . 
\end{align}
It is convenient in what follows to write the fields and their sources in
a vector notation
\begin{align}
\bscal{B}_{lm} (t,r) =
\left[
\begin{array}{c} 
h^{lm}_t\\ 
f h^{lm}_r
\end{array} 
\right] ,
\q
\bscal{V}_{lm} (t,r) = 
\left[ 
\begin{array}{c}
 f^2 P^t_{lm} \\
 -f P^r_{lm}
\end{array}
\right].	
\end{align}
The Lorenz gauge source terms are proportional to the delta function 
$\delta[r - r_p(t)]$ (in contrast to RWZ gauge where the source also has 
a $\delta'$ term), allowing the source vector to be expressed in terms of 
a time-dependent vector amplitude $\mathcal{\pmb{v}}_{lm}(t)$, 
\begin{align}
\label{eqn:oddSourceSing}
\bscal{V}_{lm} (t,r) \equiv \mathcal{\pmb{v}}_{lm} (t) \delta [r - r_p(t)].
\end{align}

The field $\bscal{B}_{lm}$ and source $\bscal{V}_{lm}$ can be expressed 
as Fourier series, analogous to Eqns.~\eqref{eqn:psiSeries} and 
\eqref{eqn:Slm},
\begin{align}
\label{eqn:oddFourierSer}
\bscal{B}_{lm}(t,r) 
&= \sum_{n=-\infty}^\infty \bscal{\ti B}_{lmn}(r) \, e^{-i \o t} , \\
\bscal{V}_{lm}(t,r) 
&= \sum_{n=-\infty}^\infty \bscal{\ti V}_{lmn}(r) \, e^{-i \o t} .
\end{align}
The FS coefficients are formally found via the integrals
\begin{align}
\bscal{\ti B}_{lmn}(r) &\equiv \frac{1}{T_r} \int_0^{T_r} dt 
\ \bscal{B}_{lm}(t,r) \, e^{i \o t},
\\
\bscal{\ti V}_{lmn}(r) &\equiv \frac{1}{T_r} \int_0^{T_r} dt 
\ \bscal{V}_{lm}(t,r) \, e^{i \o t} .
\label{eqn:Vlmn}
\end{align}
With these definitions, we henceforth in this section suppress the TD mode 
labels $lm$ and the FD labels $lmn$ whenever no confusion might arise.  
However, for clarity we attach a tilde to denote FD quantities.

In the FD, the field equations \eqref{eqn:EqnsConstrOddht} and 
\eqref{eqn:EqnsConstrOddhr} take the following form
\begin{align}
\label{eqn:oddEqsFD}
\pa_{r_*}^2 \bscal{\ti B} + \mathbf{C} \, \pa_{r_*} \bscal{\ti B} 
+ \mathbf{D} \, \bscal{\ti B} 
= \ti{\bscal{V}}.
\end{align}
The matrices $\mathbf{C}$ and $\mathbf{D}$ that couple the equations are 
given by
\begin{align}
\begin{split}
\mathbf{C} &=
\frac{2}{r^2}\left[
\begin{array}{cc} 
-M & 0 \\
0 & r-3M
\end{array} 
\right] ,
\\
\mathbf{D} &=
\l \o^2 - \frac{2 f^2 + (l+2)(l-1)f}{r^2} \r 
\mathbb{I}
\\
&\hspace{20ex} +
 \frac{2 i \o}{r^2}
\left[ 
\begin{array}{cc}
0
& - M \\
r-3M 
& 0
\end{array}
\right],
\end{split}
\end{align}
where $\mathbb{I}$ is the $2 \times 2$ identity matrix. 

The EHS method carries over to Lorenz gauge \cite{OsbuETC14,AkcaWarbBara13}.  
Four fundamental independent homogeneous solutions to 
Eqn.~\eqref{eqn:oddEqsFD} are denoted by $\bscal{\ti B}^{\pm}_i$, with 
$i=0,1$.  The $\pm$ superscript delineates causal asymptotic behavior, with 
$+$ indicating an outgoing wave at $r_*=\infty$ and $-$ indicating a 
downgoing wave at $r_*=-\infty$.  A Green function constructed from these 
arbitrarily normalized modes yields the solution to the inhomogeneous 
system 
\begin{align}
\label{eqn:oddPart}
\bscal{\ti B} =  
\bscal{\ti B}^{+}_0 c^{+}_0 +
\bscal{\ti B}^{+}_1 c^{+}_1 +
\bscal{\ti B}^{-}_0 c^{-}_0 +
\bscal{\ti B}^{-}_1 c^{-}_1 ,
\end{align}
once the $c^{\pm}_i(r)$ are determined by integrating the first-order linear 
system
\begin{align}
\label{eqn:normCDiffEq}
\mathbf{M}(r)
\left[
\begin{array}{c}
\pa_{r_*} c^{-}_0 \\
\pa_{r_*} c^{-}_1 \\
\pa_{r_*} c^{+}_0 \\
\pa_{r_*} c^{+}_1
\end{array}
\right]
=
\left[
\begin{array}{c}
\mathbf{0} \\
\ti{\bscal{V}}(r) 
\end{array}
\right] .
\end{align}
Here $\mathbf{M}$ is the ($lmn$ dependent) $4 \times 4$ Wronskian matrix
\begin{align}
&\mathbf{M}(r)  \equiv
\left[
\begin{array}{cccc}
\bscal{\ti B}^{-}_0 & \bscal{\ti B}^{-}_1 & \bscal{\ti B}^{+}_0 & 
\bscal{\ti B}^{+}_1 \\
\pa_{r_*} \bscal{\ti B}^{-}_0 & \pa_{r_*} \bscal{\ti B}^{-}_1 & 
\pa_{r_*} \bscal{\ti B}^{+}_0 & \pa_{r_*} \bscal{\ti B}^{+}_1
\end{array}
\right] ,
\end{align}
and $\mathbf{0}$ is the rank $= 2$ column vector.

Solving for the functions $c^{\pm}_i(r)$ can be avoided through use of the 
method of EHS.  Instead, as in Sec.~\ref{sec:method}, we solve for 
normalization coefficients, construct FD EHS and then TD EHS, and thus 
circumvent producing Gibbs behavior in the source region and at the particle 
location.  For the system at hand, we define the normalization constants 
$C^{\pm}_i$ as
\begin{align}
C^{+}_i \equiv c^{+}_i(r_{\text{max}}) ,
\q \q
C^{-}_i \equiv c^{-}_i(r_{\text{min}}) ,
\end{align}
and obtain them via the integrals
\begin{align}
\label{eqn:Csys}
\left[
\begin{array}{c}
-C^{-}_0 \\
-C^{-}_1 \\
C^{+}_0 \\
C^{+}_1
\end{array}
\right]
&= \int_{r_{\text{min}}}^{r_{\text{max}}} \frac{1}{f(r)} \mathbf{M}(r)^{-1} 
\left[
\begin{array}{c}
\mathbf{0} \\
\ti{\bscal{V}}(r)
\end{array}
\right]
dr .
\end{align}
In the expression above, $\mathbf{M}(r)^{-1}$ is the inverse of the Wronskian 
matrix $\mathbf{M}(r)$.  We next insert the integral expression for 
$\ti{\bscal{V}}(r)$, reverse the order of integration, and find the 
normalization coefficients with an integral over time,
\begin{align}
\label{eqn:CsysFourier}
&\left[
\begin{array}{c}
-C^{-}_0 \\
-C^{-}_1 \\
C^{+}_0 \\
C^{+}_1
\end{array}
\right]
=
\frac{1}{T_r} \int_{0}^{T_r} \frac{1}{f_p} \mathbf{M}_p^{-1}
\left[
\begin{array}{c}
\mathbf{0} \\
\pmb{v}(t)
\end{array}
\right]
e^{i\o t} dt .
\end{align}
In this last equation the script $p$ indicates time dependence via the 
mapping $r \rightarrow r_p(t)$.  With the coefficients available, the FD 
and TD EHS (respectively) are constructed
\begin{align}
\bscal{\ti B}^{\pm} (r) 
&\equiv C^{\pm}_0 \bscal{\ti B}^{\pm}_0 (r) 
+ C^{\pm}_1 \bscal{\ti B}^{\pm}_1 (r) , 
\\
\bscal{B}^\pm (t,r) 
&\equiv \sum_{n=-\infty}^\infty \bscal{\ti B}^\pm (r) \, e^{-i \o t} .
\end{align}
The solution to the system in the TD, Eqns.~\eqref{eqn:EqnsConstrOddht} and 
\eqref{eqn:EqnsConstrOddhr}, is then
\begin{align}
\label{eqn:OddSol}
\bscal{B}(t,r) &= 
\bscal{B}^+  \theta \left[ r-r_p(t)\right]
+\bscal{B}^- \theta \left[r_p(t)-r\right] .
\end{align}

The key to EHS in Lorenz gauge is solving systems like 
\eqref{eqn:CsysFourier} for the normalization coefficients.  In previous work 
\cite{AkcaWarbBara13,OsbuETC14} these equations were treated as IVPs and 
solved with ODE integration.  That numerical approach can be replaced with 
SSI to achieve spectral convergence, as we outline next.

\subsection{Spectral source integration for odd-parity normalization constants}
\label{sec:SSIlorenz}

The Lorenz gauge employment of SSI is virtually identical to RWZ gauge.  
As in Eqn.~\eqref{eqn:GFBar}, we can extract from the biperiodic source term 
$\pmb{v}(t)$ the piece that is periodic in $T_r$ by defining 
$\bar{\pmb{v}} (t) \equiv \pmb{v}(t)  e^{i m \O_{\vp}t} $.  Once substituted 
in Eqn.~\eqref{eqn:CsysFourier} we find 
\begin{align}
\label{eqn:CsysFourier2}
\left[
\begin{array}{c}
-C^{-}_0 \\
-C^{-}_1 \\
C^{+}_0 \\
C^{+}_1
\end{array}
\right]
&=
\frac{1}{T_r} \int_{0}^{T_r} 
\bm{\bar{E}}(t) \ 
e^{i n\O_r t} \ dt ,
\end{align}
where we define the vector
\begin{align}
\bm{\bar{E}}(t)
&\equiv
\frac{1}{f_p} \mathbf{M}_p^{-1}
\left[
\begin{array}{c}
\mathbf{0} \\
\bar{\pmb{v}}(t)
\end{array}
\right]	.
\end{align}
Both $\bm{\bar{E}}(t)$ and $e^{i n\O_r t}$ are periodic in $T_r$.  The 
vector $\bm{\bar{E}}(t)$, which depends on the FD labels $lmn$, is the 
equivalent of $\bar{E}^{\pm}_{lmn}(t)$ in Eqn.~\eqref{eqn:eBarDef}.

The logical steps in implementing SSI carry over from Sec.~\ref{sec:SSI}:

\bi

\item
The vector $\bm{\bar{E}}(t)$ (carrying labels $lmn$) consists of periodic, 
$C^{\infty}$ functions.  
\item
Each can be represented as a Fourier series with spectrum 
$\ti{\bm{\mathcal{E}}}_{n'}$ with $n' \rightarrow \pm\infty$.  
\item
The Fourier series spectrum can be truncated to some
$n'_{\rm min} \le n' \le n'_{\rm max}$ subject to an accuracy goal.  
\item
The approximate but very accurate truncated Fourier series is a bandlimited 
function.
\item
The Nyquist-Shannon sampling theorem implies the truncated Fourier series 
representation can itself be replaced in the TD with discrete sampling.  
\item
Sampling plus periodicity implies a discrete representation of finite 
length $N$. 
\item
Finite sampling representation in the TD implies one-to-one correspondence 
via the DFT with a FD periodic spectrum $\bm{\mathcal{E}}_{n'}$.  
\item 
The DFT spectrum within the first Nyquist minima approximates well the 
original Fourier series spectrum if $N$ is sufficiently large, allowing 
$\ti{\bm{\mathcal{E}}}_{n'} \rightarrow \bm{\mathcal{E}}_{n'}$.  

\ei
Based upon this chain of reasoning, the integral \eqref{eqn:CsysFourier2} can 
be replaced with an exponentially convergent sum
\begin{align}
\label{eqn:specSys}
\left[
\begin{array}{c}
-C^{-}_0 \\
-C^{-}_1 \\
C^{+}_0 \\
C^{+}_1
\end{array}
\right]
&= \frac{1}{N} \sum_{k=0}^{N-1} 
\bm{\bar{E}}(t_k) \, e^{i n \O_r t_k } ,
\end{align}
where again $t_k = k T_r/N$.  This is SSI for the systems of equations 
found in Lorenz gauge.

Even parity involves a larger linear system.  The matrix inversion in 
evaluating $\bm{\bar{E}}(t_k)$ at the sample points is the most expensive 
task in double precision application.  We use LU decomposition and take 
advantage of the symmetry $\mathbf{M}(t_k)=\mathbf{M}(T_r-t_k)$, so that 
LU decompositions of $\mathbf{M}$ are only necessary at $N/2 + 1$ points. 

As in Sec.~\ref{sec:SSI}, we also attain a higher rate of exponential 
convergence by switching from $t$ parametrization to $\chi$.  The adjustment 
to Eqn.~\eqref{eqn:specSys} is straightforward
\begin{align}
\left[
\begin{array}{c}
-C^{-}_0 \\
-C^{-}_1 \\
C^{+}_0 \\
C^{+}_1
\end{array}
\right]
&= 
\frac{\O_r}{N} \sum_{k=0}^{N-1} 
\frac{dt_p}{d\chi} \
\bm{\bar{E}}[t(\chi_k)] \, e^{i n \O_r t(\chi_k) },
\end{align}
where as before $\chi_k = 2 \pi k/N$.

We have implemented SSI in just this way as a modification of the Lorenz 
gauge code described in Ref.~\cite{OsbuETC14}.  SSI is particularly 
beneficial in Lorenz gauge, where a matrix must be inverted at each step 
(i.e., each function evaluation) in an integration.  It is also beneficial 
that we know precisely where the sample locations are in the source region 
before computing the inner and outer homogeneous solutions.  The previous 
method found the normalization coefficients by integrating a large simultaneous 
system of ODEs through the source region (for the even-parity field this 
tallied to integrating 144 variables simultaneously).  With prior knowledge 
of the sample locations, integration of the homogeneous solutions is 
decoupled from the SSI for the normalization coefficients. 

As we described in Sec.~\ref{sec:SSI}, the number of sample points $N$ is 
determined, iteratively if necessary, based on an error criterion.  In all 
cases we have experience with, both RWZ and Lorenz gauge, it is the source 
integration (with SSI), not the spectral integration of the orbit, that sets 
the condition on $N$.  In Lorenz gauge, the requirement on $N$ to meet a 
double precision error criterion in SSI is about a factor of 8 larger than 
required for a comparably precise orbit integration.  (With the MST code 
at a high accuracy of 200 digits, SSI requires an $N$ that is about a factor
of 2 larger than that required for comparably accurate orbit determination.)
Because SSI shrinks so markedly the computational work in finding the 
normalization coefficients, our Lorenz gauge GSF code is sped up--overall--by 
a factor of about 3 for eccentricities of order $e\simeq0.7$. 

\section{Conclusions}
\label{sec:conclusions}

We have described in this paper a new method for achieving spectral accuracy 
and computational efficiency in calculating a broad class of black hole 
perturbation and gravitational self-force problems that entail generic 
orbital motion.  This class should include most problems involving a 
point-particle description of the small compact object and use of the 
background geodesics in a frequency domain calculation (i.e., geodesic 
self-force calculations).  We have shown it applied both to the RWZ 
formalism (for individual master equations) and to Lorenz gauge (with coupled 
systems of equations) for eccentric binaries with a Schwarzschild primary.  
The method should extend to extreme-mass-ratio inspirals on Kerr as well, 
which will be addressed in subsequent work.  Called spectral source 
integration (SSI), this method provides an exponentially-convergent 
calculation of the mode normalization coefficients by replacing the more 
typically used ODE integrations in the source region.  A simple modification 
of the underlying idea is also used to integrate the equations of orbital 
motion, to provide a consistent level of accuracy in determining source 
functions in the libration region.

Use of SSI in double precision calculations will improve code speed and 
help ensure optimal accuracy.  In contrast, SSI is the \emph{sine qua non} 
for calculating eccentric binaries using (MST) analytic function 
expansions at extraordinarily high accuracies (e.g., 200 decimal places).  
No algebraically convergent ODE solver will be able to calculate 
eccentric-orbit perturbations at hundreds of decimal places of accuracy.  
Any alternative to SSI will almost certainly be a similar technique using 
some other spectral basis.  A subsequent paper will describe use of SSI in 
an MST code to uncover new terms in the post-Newtonian expansion for 
eccentric binaries well beyond known 3PN order \cite{ForsEvanHopp15}.

\acknowledgments

The authors thank Leor Barack, Scott Hughes, and Norichika Sago for helpful 
discussions and Niels Warburton for commenting on an earlier draft.  This 
work was supported in part by NSF grant PHY-1506182.  TO~acknowledges support 
from the North Carolina Space Grant's Graduate Research Assistantship Program 
and the Tom and Karen Sox Summer Research Fellowship.  EF acknowledges support 
from the Royster Society of Fellows at the University of North Carolina-Chapel 
Hill.  CRE is grateful for the hospitality of the Kavli Institute for 
Theoretical Physics at UCSB (which is supported in part by the National 
Science Foundation under Grant No. NSF PHY11-25915) and the Albert Einstein 
Institute in Golm, Germany, where part of this work was initiated.  CRE also 
acknowledges support from the Bahnson Fund at the University of North 
Carolina-Chapel Hill.  SH acknowledges support from the Albert Einstein 
Institute and also from Science Foundation Ireland under Grant 
No.~10/RFP/PHY2847.

\appendix

\section{Exact Fourier spectrum for $dt_p/d\chi$ as $p \rightarrow \infty$}
\label{sec:appendix}

Several figures in this paper have shown numerical evidence of exponential 
fall-off in the FD spectra of source functions in the perturbation and 
orbit equations.  Here we demonstrate an exact calculation of 
the Fourier spectrum in one simplified case.  Consider the source function 
$g(\chi)$ in Eqn.~\eqref{eqn:tDarwin} and make a post-Newtonian expansion
\be
\label{eqn:gOfChi}
\frac{dt_p}{d\chi} \equiv g(\chi) 
= \frac{M p^{3/2}}{(1+e \cos\chi)^2} + \mathcal{O}\left(p^{1/2}\right) .
\ee
We then focus on the leading Newtonian term and seek to find its Fourier 
spectrum.  Our derivation is similar to one found in \cite{Weid02}.  
Adopting the notation $g_N(\chi)$ for the term in question, we first 
introduce complex exponentials
\be
\label{eqn:gOfChi2}
g_N(\chi) = \frac{M p^{3/2}}
{\left[1+\frac{1}{2} e \left(e^{i\chi}+e^{-i\chi}\right)\right]^2} .
\ee
The denominator in this expression can be factored
\be
\label{eqn:gOfChi3}
g_N(\chi) = \frac{4 M p^{3/2} \sigma ^2}
{e^2  (1+\sigma e^{i\chi})^2\left(1+\sigma e^{-i\chi}\right)^2} ,
\ee
by introducing 
\be
\sigma = \frac{1}{e} \left( 1-\sqrt{1-e^2} \right) ,
\ee
which is one of the roots of the quadratic equation 
$\sigma^2 - (2/e) \sigma + 1 = 0$.  We then make a partial fractions 
decomposition of Eqn.~\eqref{eqn:gOfChi3}
\begin{align}
\label{eqn:gExp}
&g_N(\chi) =
\frac{M p^{3/2}}{e^2(\sigma^2-1)^3}\bigg[
4 \left(1+\sigma^2\right) \sigma ^2 +
\frac{4 \sigma ^2\left(\sigma^2-1\right)}{(1+\sigma e^{i\chi})^2}
\notag\\
 & \hspace{5ex}
 -\frac{8 \sigma ^4}{1+\sigma e^{i\chi}}+\frac{4 \sigma ^2\left(\sigma^2-1\right)}{\left(1+\sigma e^{-i\chi}\right)^2}
-\frac{8 \sigma ^4}{1+\sigma e^{-i\chi}}\bigg] .
\end{align}
Since it can be shown that $|\sigma| < 1$ for bound motion, each of the 
terms in Eqn.~\eqref{eqn:gExp} can be expanded in binomial or geometric
series.  The result is a Fourier series in $\chi$.  Because of the symmetry 
of $g_N(\chi)$, the expansion reduces to a cosine series
\be
g_N(\chi) = \frac{1}{2} \mathcal{G}_0 
+ \sum_{n=1}^{\infty} \mathcal{G}_n \cos\left( n\chi \right) ,
\ee
and we find that the spectrum has the form
\be
\mathcal{G}_n = \frac{8 M p^{3/2} (-1)^{n+1}}{e^2(1 - \sigma^2)^3} 
\left[ (n-1) \sigma^{n+4} - (n+1) \sigma^{n+2} \right] .
\ee
The exponential convergence of the series is evident.

\bibliography{spectralSource}

\end{document}